\documentclass[prl,twocolumn,superscriptaddress,showpacs]{revtex4}
\usepackage[italian,english]{babel}

\usepackage{graphicx}
\newcommand{\lf}{\left}
\newcommand{\rg}{\right}
\newcommand{\be}{\begin{equation}}
\newcommand{\ee}{\end{equation}}
\newcommand{\bea}{\begin{eqnarray}}
\newcommand{\eea}{\end{eqnarray}}
\newcommand{\ba}{\begin{array}}
\newcommand{\ea}{\end{array}}
\newcommand{\bd}{\begin{displaymath}}
\newcommand{\ed}{\end{displaymath}}
\renewcommand{\a}{\alpha}
\renewcommand{\b}{\beta}
\newcommand{\g}{\gamma}
\newcommand{\G}{\Gamma}

\renewcommand{\l}{\lambda}
\newcommand{\n}{\nu}

\renewcommand{\O}{\Omega}

\newcommand{\hf}{\frac{1}{2}}

\begin{document}

\title{Non-abelian superconducting pumps}

\author{Valentina Brosco}
\affiliation {Institut f\"ur Theoretische Festk\"orperphysik and DFG
Center for Functional Nanostructures (CFN),
        Universit\"at Karlsruhe, 76128 Karlsruhe, Germany}
\affiliation{NEST-CNR-INFM \& Dipartimento di Fisica, Universit\'a
di Pisa,
        largo E. Fermi, I-56100 Pisa, Italy}
\author{Rosario Fazio}
\affiliation{International School for Advanced Studies (SISSA)
        via  Beirut 2-4,  I-34014, Trieste - Italy}
\affiliation{NEST-CNR-INFM \& Scuola Normale Superiore, piazza dei
        Cavalieri 7, I-56126 Pisa, Italy}
\author{F.W.J. Hekking}
\affiliation{LPMMC, CNRS \& Universit\'e Joseph Fourier, BP 166,
        38042 Grenoble CEDEX 9, France}
\author{Alain Joye}
\affiliation{LPMMC, CNRS \& Universit\'e Joseph Fourier, BP 166,
        38042 Grenoble CEDEX 9, France}
\affiliation{Institut Fourier, BP 74, 38402 Saint-Martin d'H\`eres,
France}

\begin{abstract}
Cooper pair pumping is a coherent process. We derive a general
expression for the adiabatic pumped charge in superconducting
nanocircuits in the presence of level degeneracy and relate it to
non-Abelian holonomies of Wilczek and Zee. We discuss an
experimental system where the non-Abelian structure of the adiabatic
evolution manifests in the pumped charge.
\end{abstract}

\pacs{}

\maketitle

\begin{figure}[t!]
\begin{center}
\includegraphics[scale=0.5, width=0.5\textwidth]{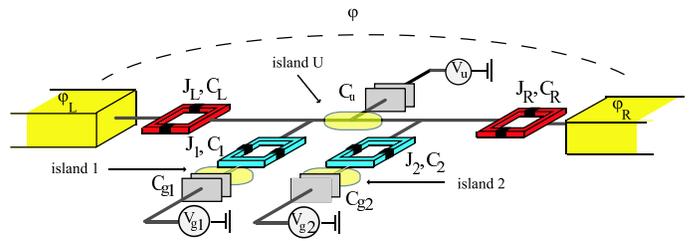}
\caption{Non-abelian superconducting pump. The gate voltages are
kept fixed during the cycle, their values are chosen in order to
have a doubly degenerate spectrum. The only pumping parameters are
the fluxes through the four loops.} \label{pump}
\end{center}
\end{figure}

If the Hamiltonian of a quantum system depends adiabatically and
cyclically on time via some external parameters the wavefunction,
for the system initially prepared in an energy eigenstate, after a
cycle returns back to its initial state up to a phase, that in
addition to the dynamical contribution, has a component of geometric
nature~\cite{berry}. It  depends only on the shape of the path
covered in the parameters space and on the structure of the Hilbert
space of the quantum system. If the adiabatic evolution takes place
in a degenerate energy eigenspace the cyclic evolution
%
%
 leads to a superposition of the degenerate
eigenstates and the geometric transformation acquires a non-Abelian
structure~\cite{wilczek}. Simon~\cite{simon} and Wilczek and
Zee~\cite{wilczek} showed that this mapping plays the role of the
holonomy of the gauge theories. Holonomies naturally emerge in the
description of the dynamics of simple quantum systems due to the
parallel transport dictated by the Schroedinger
equation~\cite{shapere,Bohm} which constraints the overlap between
the wavefunctions of the system at successive times to be real and
positive. The holonomy group is the group of the transformations
generated by the parallel transport along closed paths on the
parameter space.\\
\indent Geometric effects appear naturally in  {\em adiabatic
quantum pumping}: in a mesoscopic conductor a dc charge current can
be obtained, in the absence of applied voltages, by cycling in time
two parameters which characterize the system~\cite{thouless}. In the
scattering approach to transport the pumped charge per cycle can be
expressed in terms of derivatives of the scattering amplitudes with
respect to the pumping parameters~\cite{buttiker}, the Brouwer
formula~\cite{brouwer}. Its relation to geometric phases has been
elucidated in Refs.\cite{avron,mckenzie}. If only superconducting
leads are present
pumping is due to the adiabatic transport of Cooper pairs. Besides
the dependence of the pumped charge on the cycle, in this case there
is a dependence on the superconducting phase difference(s) (the
overall process is coherent). Cooper pair pumping has been studied
both in the limit of transparent
interfaces~\cite{governale,footnote} and in the Coulomb Blockade
regime~\cite{geerligs,pekola,niskanen}.  A connection between Berry
phase and pumped charge has been established also in this
case~\cite{aunola,governale,mottonen} thus opening the possibility
to detect geometric phases in superconducting
circuits~\cite{mottonen,falci}. An experiment of this kind has been
successfully performed recently in Ref.~\cite{mottonen07} thus
paving the way to holonomic quantum computation~\cite{jones,zanardi}
with superconducting nanodevices.\\
\indent In this Letter we study Cooper pair pumping in
superconducting circuits in the regime of Coulomb blockade. The new
feature we consider here is the possibility to pump in a degenerate
subspace.
We derive an expression for the pumped charge in the presence of a
degenerate spectrum and relate it to the non-Abelian connection of
Wilczek and Zee. We propose a superconducting network where this
relation can be tested and discuss two clear signatures of
non-abelian holonomies. First, under appropriate conditions, the
pumped charge per cycle is quantized. Second the pumped charge
depends {\em both} on the cycle and on the point where the cycle
starts. If tested experimentally this would be
a clear proof of the non-Abelian nature of pumping.\\
The possibility to generate non-abelian holonomies in
superconducting circuits has been studied previously ~\cite{faoro}.
Ref.\cite{faoro} dealt with the problem of implementing a holonomic
quantum computer with Josephson circuits. In that work the  authors
show that a cyclic adiabatic change of parameters in a closed system
may lead to a final state having a different charge distribution
than the initial one. In that context the word  "charge  pumping"
was used to describe such process even if the charge cannot be  
really pumped in or out the system. In the present work  we study 
charge transport between two external reservoirs connected to the 
circuit via Josephson junctions, the word "pumping" has thus a
different meaning.
In order to keep the presentation transparent we discuss an
idealized situation, in the conclusions we discuss the various
problems that may occur in experiments. The Cooper pair pump
consists of a Josephson network connected through Josephson
junctions to two superconducting leads. An example (at this stage
the discussion is general) is presented in Fig.~\ref{pump}. The two
superconducting electrodes are kept at a finite phase difference
$\varphi = \varphi_L - \varphi_R$ where $\varphi_{R/L}$ is the phase
of the superconducting order parameter of the right/left lead. The
Cooper pair pump is operated by changing adiabatically in time some
external parameters such as gate voltages or magnetic fluxes. We
will label this set of external parameters by the vector $\vec
\l(t)=\{V_{gi}(t),\Phi_{i}(t)\}$. The Hamiltonian of the pump
depends on the superconducting phase of each island of the network
$\varphi_i$ ($i=1,\ldots N$), on its conjugate, the charge on each
island $n_i$, on the phase difference across the pump, $\varphi$,
and on all the external parameters:
$H(t)=H\big[\varphi_1,...,\varphi_N; n_1,...,n_N;\vec\l(t),
\varphi\big]$. The state of the system is denoted by
$\lf|\Psi(t)\rg>$  and
 $\lf|\Psi(t)\rg>=\big|\Psi\big[t,\vec\l(t),\varphi\big]\big>$.
By changing the control parameters in time, a charge $Q^{(tr)}$ will
be transferred in the circuit. The transferred charge after a period
$T$ can be obtained by integrating the charge imbalance between the
outer capacitors that connect the network to the leads
\be \label{qdef} Q^{(tr)}=-2ie\int_0^T\partial_{t'}
\lf[\lf<\Psi(t')\rg|\partial_{\varphi}\lf|\Psi(t')\rg>\rg]dt' \;\; .
\ee
This definition of the transferred charge may be derived from the
time integral of the current operator,
$Q^{(tr)}=-\frac{2e}{\hbar}\int_0^T
\lf<\Psi(t')\rg|\lf(\partial_{\varphi}H(t)\rg)\lf|\Psi(t')\rg>dt'$,
and the Schr\"odinger equation. In the definition of the transferred
charge $\partial_{\varphi}$ is not a quantum operator but a simple
derivative respect to a classical parameter (the phase difference
between the two  electrodes). In this Letter we generalize the
results obtained so far  relating Cooper pair pumping to geometric
phases allowing the spectrum of $H(t)$ to be degenerate. We assume
that for all $\vec \l$ in the parameter space a degenerate energy
eigenspace $\mathcal{H}_n\big(\vec \l\big)$ exists of constant
dimension, $D_n$, corresponding to the eigenvalue
$E_n\big(\vec\l\big)$. The control parameters are varied in time
adiabatically, i.e. $\vec \lambda=\vec \lambda(t/T)$. It is
convenient to introduce for all $t\in[0,T]$ a basis,
$\lf\{\psi_{n\a}(t)\rg\}$ $\a\in[1..D_n]$, of the degenerate
subspace, $\mathcal{H}_n\big(\vec \l\big)$, formed by the
instantaneous eigenstates of the hamiltonian, $H(t)$:
$H(t)\lf|\psi_{k\a}(t)\rg>=E_k(t)\lf|\psi_{k\a}(t)\rg> $. As
discussed by Wilczek and Zee~\cite{wilczek}, if initially the state
of the system is in one of the degenerate eigenstates
$\lf|\Psi(t=0)\rg>=\big|\psi_{n\n}[\vec\l(0)]\big>\in\mathcal{H}_n\big(\vec\l(0)\big)$,
then after a cyclic evolution \be \label{psi}
\lf|\Psi(T)\rg>=\lf[U_n(T)\rg]_{\a\n}\big|\psi_{n\a}(\vec\l(0))\big>
+ {\cal O}(1/T)\; . \ee The  $D_n\times D_n$ operator $U_n(t)$ can
be written as
 \be
\label{Ukab} U_n(T)=e^{-\frac{i}{\hbar}\int_0^T E_n(t)}\mathcal{T}e^
{\lf[-\int_0^T\Gamma_n(t)dt\rg]}. \ee Here, $\mathcal{T}$ denotes
the time-ordering and  the connection $\Gamma_n(t)$ is given by
$\lf[\Gamma_n(t)\rg]_{\a\b}
=\lf<\psi_{n\a}(t)\rg|\lf.\dot\psi_{n\b}(t)\rg> $. The relation
between the transferred charge and the non-Abelian holonomy can be
obtained by substituting (\ref{psi}) in (\ref{qdef}) \cite{n1}. We
assume that  the system is prepared  at $t=0$ in a linear
superposition of the  degenerate eigenstates,
$\big|\,\Psi_{in}\big>=\sum_{\g}c_\g\big|\,\psi_{n\g}[\vec\l(0)]\big>$;
during a cycle of duration $T$ the total transferred charge then
reads
\begin{equation}\label{qtr}
Q^{(tr)}=\sum_{\g\g'}c^*_\g c_{\g'} \hat Q^n_{\g\g'},
\end{equation}
where the charge matrix $\hat Q^n_{\g\g'}$ is given by
\begin{widetext}
\begin{equation}
\hat Q^n_{\g\g'} =-\frac{2e}{\hbar}\int_0^T\! \lf\{\partial_\varphi
E_n\delta_{\g\g'}-i\hbar\sum_{\a\b}\lf[[U^\dag_n]_{\g\a}\big(\partial_\varphi\lf[\G_n\rg]_{\a\b}\big)[U_n]_{\b\n}
-\,\partial_t\!\lf([U^\dag_n]_{\n\a}\lf<\psi_{n\a}\rg|\partial_{\varphi}\lf|\psi_{n\b}\rg>[U_n]_{\b\g'}\rg)\rg]\rg\}\,dt
+ {\cal O}(1/T). \label{qnn}
\end{equation}
\end{widetext}
The first term in the r.h.s.~is the supercurrent contribution to the
transferred charge. The second and third terms are of geometrical
nature and describe pumping. 
If the pumping
occurs through a non-degenerate level, $D_n=1$, we recover the
abelian result ~\cite{aunola,mottonen,footnote2}. The measurement of
non abelian holonomies is a non-trivial task, in particular the non
commutativity of the theory has revealed not easy to
detect~\cite{Bohm}. Non-abelian contributions to pumping introduce
{\em qualitatively new} effects that can be verified experimentally.
We will highlight these aspects by analyzing a specific case that
can be realized experimentally.

A possible experimental realization of a non-abelian pump is shown
in Fig.\ref{pump}. It is a  three island pump with four symmetric
Josephson SQUID loops: two inner loops with capacitances $C_1$ and
$C_2$ and two outer loops with capacitances $C_L$ and $C_R$. The
outer loops connect the network to the superconducting electrodes
which are kept at a constant phase difference $\varphi$ (from now on
for simplicity we fix $\varphi_L = \varphi$ and $\varphi_R=0$). The
charging configuration of the system can be controlled externally
modulating three gate voltages, $V_u$, $V_{g1}$ and $V_{g2}$
connected to the islands via the respective gate capacitances,
$C_u$, $C_{g1}$ and $C_{g2}$. The effective Josephson couplings,
$J_L,\,J_R,\,J_1,\,J_2$, can be tuned independently varying the
magnetic fluxes through each loop. 
%
All the Josephson coupling energies are much smaller than the
charging energy of the system, $E_C$. The charge states are
indicated as $\lf|n_u,n_1,n_2\rg>$, $n_i$ being the excess charge on
the $i$-th island in units of $2e$. The realization of the
degenerate subspace requires additional constraints on the gate
voltages. We take the gate capacitances to be small compared to the
Josephson capacitances, i.e. $C_u\sim C_{gi} \ll C_i\sim C_I$ with
$i=1,2$ and $I=L,R$, and we assume that the gate charge
$n_{gu}=C_uV_{gu}/2e$ is in the range $\hf<
n_{gu}<\frac{3}{2}+\frac{C_1+C_2}{C_L+C_R}$ while the two gate
charge $n_{gi}=C_{gi}V_{gi}/2e$ satisfy the condition
$n_{gi}=\hf\lf(1+\frac{C_i}{C_T}(2n_{gu}-1)\rg)$  with $i=1,2$ and
$C_T=C_1+C_2+C_L+C_R$. Under these conditions only four charge
states are relevant to the dynamics at low temperatures:
$\lf|0,0,0\rg>,\,\lf|0,1,0\rg>,\lf|0,0,1\rg>,\,\lf|1,0,0\rg>$ (all
other charge states are at a much higher energy $\sim E_C$).
Moreover the charge states
$\lf|0,0,0\rg>,\,\lf|0,1,0\rg>,\lf|0,0,1\rg>$ are degenerate while
the charge state $\lf|1,0,0\rg>$, corresponding to the configuration
in which there is one excess charge on the island $U$, has higher
energy. In this restricted Hilbert space the effective hamiltonian
of the pump can be written as \bea H &=& E_u |1,0,0\rangle \langle
1,0,0| + \left[ J_{\rm eff}(\varphi) |1,0,0\rangle \langle 0,0,0|
\right.
\nonumber \\
& +& \left. J_1  |1,0,0\rangle \langle 0,1,0|
  + J_2  |1,0,0\rangle \langle 0,0,1| +\mbox{h.c.} \right]
\label{ham} \eea where $J_{\rm eff}(\varphi)=(J_Le^{i\varphi}+J_R)$.
We set to zero the electrostatic energy of the degenerate charge
configurations and we denoted with $E_u$ the charging energy of the
state $\lf|1,0,0\rg>$, $E_u=\frac{2e^2C_1(2n_{g1}-1)}{(C_L+C_R)^2}$.
Given the capacitances, our choice of the gate voltages guarantees
that the charging hamiltonian is symmetric under the simultaneous
exchange of the three charge states and of the three couplings
$J_{\rm eff}$, $J_1$ and  $J_2$. This fact leads to a two
dimensional degenerate subspace for any value of the couplings.
Hamiltonian~(\ref{ham}) has been discussed previously in the context
of adiabatic passage techniques in a quantum optics~\cite{unanyan}
and superconducting nanocircuits~\cite{siewert} and for  holonomic
quantum computation as demonstrated in~\cite{duan}.  The crucial
point 
here is that Eq.~(\ref{ham}) arises from a
Josephson network \emph{in the presence of superconducting
electrodes}. This is why we are able to relate pumping to
holonomies. The hamiltonian has three distinct eigenvalues:
$E_0=1/2\,\lf[E_u-(E_u^2+4(J_0^2+J_1^2+J_2^2))^{1/2}\rg]$, $E_1=0$
and $E_2=1/2\,\lf[E_u+(E_u^2+4(J_0^2+J_1^2+J_2^2))^{1/2}\rg]$. In
the previous definitions we set
$J_0=\lf(J_L^2+J_R^2+2J_LJ_R\cos\varphi\rg)^{1/2}$. The eigenvalue
$E_1$ remains zero and it is doubly degenerate for any value of the
Josephson couplings not all zero. An orthonormal basis for the
two-dimensional degenerate subspace corresponding to the eigenvalue
$E_1$ is given by : $\lf|\psi_{11}\rg> =
N_{11}\,(J_2\lf|0,1,0\rg>-J_1\lf|0,0,1\rg>) $ and $\lf|\psi_{12}\rg>
=N_{12}\,\lf[(J_1^2+J_2^2)\lf|0,0,0\rg>-J_{\rm eff}(\varphi)
\lf(J_1\lf|0,1,0\rg> +J_2\lf|0,0,1\rg>\rg)\rg]$ where $N_{11}$ and
$N_{12}$ are normalization factors.

\begin{figure}[b!]
\begin{center}
\includegraphics[scale=0.5]{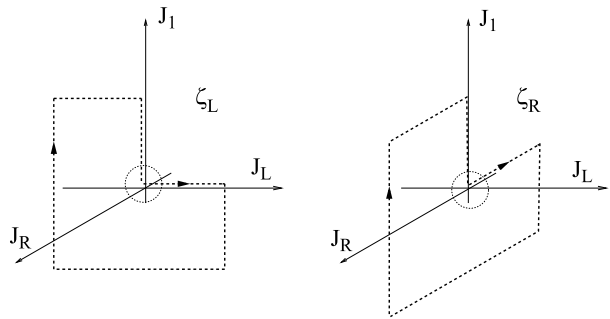}
\caption{By varying the Josephson couplings through the cycles
$\zeta_L$ (left) and then $\zeta_R$ (right)  one Cooper pair is
transferred through the pump.} \label{cycle}
\end{center}
\end{figure}

In the following we will show that: (i)   The pumped charge may be
quantized; (ii) By composing two cycles in the parameter space, the
pumped charge
     depends on the order in which the two cycles are followed.
(iii) Considering the pumped charge as a function of time, $Q=Q(t)$,
the period of $Q(t)$ may be a multiple of the period of the
parameters cycle. The  second and the third point are due to the
non-abelian character of the adiabatic evolution.

(i) We first discuss the quantization of the pumped charge. To this
end the cycle can be divided in three steps. First, we prepare the
system in the charge state $\lf|0,0,0\rg>$ and we set all the
couplings to zero except for $J_2$ which is kept constant and
positive during the whole cycle.  For these initial values of the
couplings the two degenerate eigenstates are also charge
eigenstates: $\lf|\psi_{11}(t=0)\rg>=\lf|0,1,0\rg>$,
$\lf|\psi_{12}(t=0)\rg>=\lf|0,0,0\rg>$; so  the system is initially
in the  eigenstate $\lf|\psi_{12}\rg>$. Second, we perform a $\pi/2$
rotation in the degenerate subspace by manipulating adiabatically
and cyclically $J_L$ and $J_1$ and keeping $J_R$ to zero. In this
phase one charge enters the pump from the left reservoir and the
state vector of the system undergoes the
 transformation:
$\lf|0,0,0\rg>\rightarrow e^{i\varphi}\lf|0,1,0\rg>$. Third, we let
the charge out of the circuit with  another $\pi/2$ rotation in the
degenerate subspace. During this third phase we manipulate $J_R$ and
$J_1$ and we keep $J_L$ to zero. The state of the system undergoes
the transformation $ e^{i\varphi}\lf|0,1,0\rg>\rightarrow
e^{i\varphi}\lf|0,0,0\rg> $. How is it possible to realize the
$\pi/2$ rotations? One can show that by manipulating adiabatically
$J_1$,$J_2$ and $J_L$ or $J_1$,$J_2$ and $J_R$ and keeping
respectively $J_R$ or $J_L$ zero the holonomy reads
\be U_{L/R}^{(\zeta)}=\left(%
\begin{array}{cc}
  \cos\O_\zeta & e^{i\varphi_{L/R}}\sin\O_\zeta \\
  -e^{-i\varphi_{L/R}}\sin\O_\zeta & \cos\O_\zeta \\
\end{array}%
\right),\ee where $\O_\zeta = \oint_{\zeta} \langle \psi_{11}|
\dot{\psi}_{12}\rangle $.
%
%
The required rotations are realized whenever $\O_\zeta=\pi/2$, for
example by means of the cycles $\zeta_L$ and $\zeta_R$ shown in
Fig.~\ref{cycle}. By substituting the evolution $U$, the connection
$\Gamma$  and the matrix elements of $\lf<\psi_{n\a}|
\partial_{\varphi} |\psi_{n\b}\rg>$ for the cycles of Fig.~\ref{cycle} in
Eq.~(\ref{qnn}) we find $ Q^{(tr)} = -2e. $ For the present cycle,
the charge pumped starting from the eigenstate $\lf|\psi_{11}\rg>$
has opposite sign to the charge pumped starting from the eigenstate
$\lf|\psi_{12}\rg>$.
\begin{figure}
\includegraphics[scale=0.5]{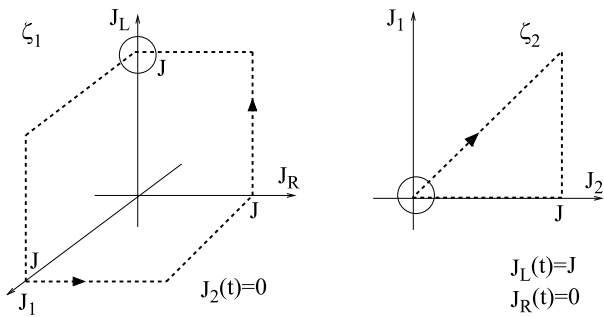}\caption{Non commuting cycles in the
parameters space. Along $\zeta_1$,  $J_2$ is set to zero while along
$\zeta_2$,  both $J_L$ and $J_R$ are fixed. The circles indicate the
initial points of each cycle.} \label{noncom}
\end{figure}
(ii) The non-abelian nature of the evolution has measurable effects
on the pumped charge. We consider the cycles depicted in
Fig.~\ref{noncom} and determine the pumped charge in the two cases
in which the pumping cycle is either performed by first going over
$\zeta_1$ and then $\zeta_2$ or in the reverse order. In the abelian
case considered in Refs.~\cite{pekola,aunola,mottonen} the two
situations are equivalent. In our case the order matters, i.e. there
are examples where $ Q^{(tr)}_{\zeta_1\zeta_2} \ne
Q^{(tr)}_{\zeta_2\zeta_1}$. A specific example is provided by the
pumping cycle obtained by performing the two loops presented in
Fig.~\ref{cycle}. Assuming that the initial state is the state
$|0,1,0\rangle$, the pumped charge in the two cases is $
Q^{(tr)}_{\zeta_1\zeta_2}= \frac{e}{2}$ and
$Q^{(tr)}_{\zeta_2\zeta_1}= e$. The pumped charge {\em differs} in
the two cases.

(iii) Another manifestation of the non-Abelian nature of pumping is
that after a cyclic evolution of the external parameters the state
does not necessarily go to the initial state (see Eq.~(\ref{psi})).
One can therefore design paths in parameter space such that after
$N$ cycles the system returns, up to a phase, to its initial state.
In this situation the pumped charge per cycle will not be constant
in each cycle but it will have a period which is $N\,T$, $T$ being
the period of the elementary cycle. In fact, assume that a certain
cycle $\bar \zeta$ is performed consecutively N times starting with
the system in the state $|\Psi_{in}\big>$ defined before. At the
beginning of the N-th cycle the state will be
$\lf(U_{\bar\zeta}\rg)^{N-1}\big|\Psi_{in}\big>$ then, as one can
easily show using Eq. (\ref{qtr}), the total transferred charge will
be given by: $Q_{{\rm Nth\,cycle}}^{tr}= \sum_{\g\g'}c^*_\g
c_{\g'}\lf[\lf(U^\dag_{\bar\zeta}\rg)^{N-1} \hat
Q^n\lf(U_{\bar\zeta}\rg)^{N-1}\rg]_{\g\g'}$. Eventually, when
$U_{\bar\zeta}^{N-1}$ is proportional to  the identity operator, the
transferred charge will be  periodic with period $N\,T$.

The situation we discussed so far is ideal. Several important issues
have to be considered in a realistic situation. First of all, it
would be desirable to pump through the ground state;  here the
degenerate subspace is the first excited level, which is sensitive
to relaxation effects. We do not think this is a serious problem,
though: we showed how the state can be accessed and moreover the
real bottleneck is the decoherence time and not the relaxation time,
which typically is much longer. Indeed, the effect of decoherence on
coherent pumping appears to be a more fundamental issue since the
presence of an external bath may, in addition to relaxation, lift
part of the degeneracy which is crucial for non-Abelian pumping. In
addition, the degeneracy may be lifted because of the unavoidable
static imperfections in the network. We do not expect degeneracy
lifting to prevent the observation of non-Abelian effects on
pumping, it just imposes a constraint on the duration of the cycle:
$T$ should be shorter than $\mbox{min}\{\hbar/\Delta
E,\hbar/\tau_{\phi}\}$ where $\Delta E$ is the small splitting
arising from the non perfect degeneracy of the levels involved and
$\tau_{\phi}$ the decoherence time.

This work has been supported by MIUR-PRIN, Institut universitaire de
France and EC-Eurosqip. The support of the Lewiner Institute for
Theoretical Physics at the  Technion is acknowledged.

\end{document}